# Free-Choice Petri Nets without frozen tokens and Bipolar Synchronization Systems

Joachim Wehler

*Ludwig-Maximilians-Universität München, Germany*

joachim.wehler@gmx.net

**Abstract:** *Bipolar synchronization systems (BP-systems) constitute a class of coloured Petri nets, well suited for modeling the control flow of discrete, dynamical systems. Every BP-system has an underlying ordinary Petri net, which is a T-system. Moreover, it has a second ordinary net attached, which is a free-choice system. We prove that a BP-system is safe and live if the T-system and the free-choice system are safe and live and if the free-choice system has no frozen tokens. This result is the converse of a theorem of Genrich and Thiagarajan and proves an old conjecture. For the proof we introduce the concept of a morphism between coloured Petri nets as a means to compare different Petri-nets. Then we apply the classical theory of free-choice systems.*

# Introduction

Bipolar synchronization systems (BP–systems) constitute a class of coloured Petri nets, well suited for modeling the control flow of discrete distributed dynamical systems. BP-systems have been introduced in 1984 by Genrich and Thiagarajan [GT1984].

BP–systems have two token colours, high-tokens and low-tokens, and they have coloured transitions with firing modes depending on the combination of high- and low-tokens at their preplaces. As a consequence, a transition decides not only about activating a subsequent activity but also about skipping it. The flow of high-tokens shows the pattern of activation, the flow of low-tokens the pattern of skipping activities. The firing modes of a given transition obey either an AND-rule or a XOR-rule.

Genrich and Thiagarajan observed that the flow of high-tokens of a BP-system induces the token flow of a corresponding free-choice system. We call it the high-system of the BP-system. Abstracting from the colours of a BP-system leads to a second ordinary Petri net. This T-system yet keeps the net structure of places, transitions and directed arcs. We call it the skeleton of the BP-system. The forgetting about the colours is formalized by a canonical morphism from the BP-system to its skeleton. Due to this morphism the safeness of a BP-system follows from the safeness of its skeleton. Conversely, liveness and safeness of a BP-system imply the analogous properties of its skeleton, thanks to a lifting lemma for the Petri net morphism.

Genrich and Thiagarajan already proved, that the high-system of a safe and live BP-system is safe and live itself. Moreover the high-system has no frozen tokens. Both results follow from the lifting lemma. The new result of the present paper proves the converse of the theorem of Genrich and Thiagarajan. Our main result (Theorem 4.6):
A BP-system is safe and live iff its high-system is safe and live without frozen tokens and its skeleton is safe and live.
For the proof we conclude from the lifting lemma, that deadlock-freeness is sufficient for the liveness of the BP-system. This result has also already been shown by Genrich and Thiagarajan. But liveness and safeness of high-system and skeleton do not suffice to exclude a deadlock of the BP-system. Therefore we intensify the concept of a deadlock to the stronger concept of a deadlock-configuration. It consists of an alternating series of closing XOR- and AND-transitions. Firing the AND-transition in the high-system presupposes firing the XOR-transition, yet firing the XOR-transition in the skeleton presupposes firing the AND-transition. Hence the transitions in the BP-system block each other. We prove, that every dead BP-system has a deadlock-configuration, if its high-system and skeleton are safe and live. On the other hand, any deadlock-configuration is excluded by the absence of frozen tokens.
The essential means for proving our main theorem are: The high-net belongs to a class of restricted free-choice nets where well-formedness can be characterized by the absence of certain handles on elementary circuits. Using circuits allows us to carry a common type of reasoning from T-systems to the high-net of a BP-system. Moreover, we use a theorem about the activation of T-components in safe and live free-choice systems. Both types of arguments were not at disposal of Genrich and Thiagarajan in 1984, they were developed afterwards by Best, Desel, Esparza and Silva. A third input for our proof is the simple observation, that an activated T-component in a Petri net without frozen tokens must already contain all tokens.

# 1 Free-choice Petri nets

In this chapter we recall some well-known results from the classical theory of free-choice systems. Our basic reference is [DE1995]. Free-choice systems will play an important role when attaching certain ordinary Petri nets to a bipolar synchronization system later on.
A P-component of a free-choice system with exactly one token is named a *basic component*. An elementary circuit with exactly one token is called a *basic circuit*. A T-component of a free-choice system is *activated* at a reachable marking, iff it is live with respect to the restricted marking. For a subnet $A$ of a given net, we denote by $cl(A)$ the subnet of the given net generated by all clusters with a place in $A$. A free-choice net is named *restricted free-choice net*, iff for every place $p$ holds

$$post(p) > 1 \Rightarrow pre\,(post\,(p)\,) = p$$

A marked restricted free-choice net is called *restricted free-choice system*.

## 1.1 Theorem (*Activation of T-components*)

Every $T$ -component $N_T$ of a live and bounded free-choice system can be activated by a reachable marking. There exists an activation, which does not fire any transition from $cl(N_T)$.

The proof from [DE1995], Theor. 5.20, also demonstrates the stronger version of Theorem 1.1. The following Corollary 1.2 is well-known, cf. [DE1995], Lemma 9.15 for its proof.

## 1.2 Corollary *(Embedding an elementary circuit into components)*

Any elementary circuit of a well-formed free-choice net is contained in the intersection of a $P$-component with a $T$-component.

A token in a Petri net is frozen at a given place, iff the current marking activates an infinite occurrence sequence, which never moves the token.

## 1.3 Definition *(Frozen tokens)*

An ordinary Petri net $(N, \mu_0)$ has no *frozen tokens*, iff for every reachable marking $\mu$ holds: For every strictly less marking $\nu < \mu$ the Petri net $(N, \nu)$ has no infinite occurrence sequence.

For a free-choice system the absence of frozen tokens is a structural property.

## 1.4 Definition *(Structural free from blocking)*

A free-choice net is *structural free from blocking*, iff every $P$-component intersects every $T$-component in a non-empty set.

## 1.5 Lemma *(Frozen tokens)*

A safe and live free-choice system has no frozen tokens, iff the underlying net is structural free from blocking.
**Proof**. Cf. [BD1990], Theor. 6.2.S

Any strongly connected T-system is structural free from blocking. Hence a safe and live T-system has no frozen tokens.
Because the non-empty intersection of a P-component with a T-component in a free-choice net is a set of disjoint non-empty elementary circuits [Des1992], we obtain as a corollary:

## 1.6 Corollary *(Intersection of components)*

In a safe and live free-choice system without frozen tokens the intersection of a basic component with a $T$-component is a non-empty elementary circuit. In the case of an activated $T$-component the intersection is even a basic circuit.

For the class of restricted free-choice nets – but not for free-choice nets in general – there exists a characterization of well-formedness in terms of certain paths, named resp. handles and bridges [Des1986], [ES1990].

## 1.7 Theorem *(Well-formedness of restricted free-choice nets)*

A restricted free-choice net is well-formed iff it is strongly connected, no elementary circuit has a TP-handle, and every PT-handle of an elementary circuit has a TP-bridge.

## 1.8 Definition *(Blocking marking)*

A marking of a free-choice net is called a *blocking marking* of a cluster of the net, if the marking activates every transition from the cluster but no other transition.

## 1.9 Lemma *(Blocking markings in the absence of frozen tokens)*

Any cluster of a safe and live free-choice system without frozen tokens has a blocking marking, which is uniquely determined and can be reached from every reachable marking.

**Proof**. Denote the given free-choice system by $FCS = (N, \mu_0)$. Because $FCS$ has no frozen tokens, for any cluster of $N$ and at every reachable marking there exists an occurrence sequence, the firing of which creates a blocking marking of the given cluster.

The second statement about the uniqueness of a blocking marking will be reduced to the special case of T-systems, where it is well-known. Claim: Any T-component of $N$ , which contains at least one token of a blocking marking $\mu_{block}$ must contain all other tokens of the blocking marking too. For the proof we note, that any T-component, which is activated at a reachable marking $\mu$ , must contain all tokens from $\mu$ , because $FCS$ has no frozen tokens. Now consider a T-component $N_T$ with a place $p \in N_T$ marked at $\mu_{block}$ . If $\mu_{block}$ would mark a second place $q \in N - N_T$ in the complement of $N_T$ , one could not activate $N_T$ at $\mu_{block}$ without firing a transition from $post(q) \subset cl(N_T)$. This contradicts Theorem 1.1 and proves the claim. Hence a blocking marking of $FCS$ is also a blocking marking of a suitable, activated T-component $N_T$ . According to ([GT1984], Theor. 1.15) any blocking marking of a fixed cluster within a save and life T-system is uniquely determined. As a consequence any blocking marking of a fixed cluster in $FCS$ is also uniquely determined, q. e. d.

The result of Lemma 1.9 has been generalized to arbitrary bounded and live free-choice systems by Gaujal, Haar and Mairesse, cf. Remark 1.10. We will not apply their result, because its proof is much deeper than the proof of Lemma 1.9, which presupposes the absence of frozen tokens.

## 1.10 Remark *(Existence and uniqueness of blocking markings)*

Every cluster of a bounded and live free-choice system has a unique blocking marking. From every reachable marking the blocking marking can be obtained by firing an occurrence sequence, which does not contain an arbitrary, but fixed transition from the cluster.

The authors prove this theorem in ([GHM2003], Theor. 3.1) for restricted free-choice systems and clusters with only one transition (non-conflicting transitions). In section 6.1 of their work they apply a simple transformation to reduce the case of a free-choice system to the case of a restricted free-choice system. The case of a general cluster reduces to the case of a non-conflicting transition by another simple transformation.

# 2 BP-Systems and their derived ordinary Petri nets

A BP-system is a coloured Petri net. Hence we first recall the definition of coloured Petri nets from [Jen1982], cf. also [Jen1992]. Subsequently we develop the concept of a morphism between coloured Petri nets. Therefore we have to recast the definition of coloured Petri nets, separating their topological properties from their algebraic features.

## 2.1 Definition *(Coloured Petri net)*

i) A *coloured net* is a tuple

$$CN = \left( P, T, C(p)_{p\in P}, B(t)_{t\in T}, \left(w^{-/+}{}_{t,p}\right)_{(t,p)\in T\times P} \right)$$

with:

- Two non-empty, finite disjoint sets $P$ (*places*) and $T$ (*transitions*)
- a family of non-empty, finite sets $C(p), (\textit{token colours}), p \in P$,
- a family of non-empty, finite sets $B(t), (\textit{firing modes}), t \in T$,
- and two families, the *negative* resp. *positive incidence-functions*,

$$w^{-}{}_{t,p}, w^{+}{}_{t,p} : B(t)_N \longrightarrow C(p)_N, (t,p) \in T \times P,$$

of $N$ -linear maps between the free commutative monoids $B(t)_N$ and $C(p)_N$ over the set resp. $B(t)$ and $C(p)$.

ii) A *coloured Petri net* is a pair $PN = (CN, \mu)$ with a coloured net $CN$ and a marking $\mu$ of $CN$.

A pair $(p,c)$, $p \in P, c \in C(p)$, is called a *token element* of $CN$, while a *firing element* of $CN$ is a pair $(t,b)$, $t \in T, b \in B(t)$. A BP-system is a coloured Petri net, which captures the logical operations AND and XOR.

## 2.2 Definition *(BP-system)*

i) A *bipolar synchronization graph* (BP-graph) is a coloured net $BPG = \left( P, T, C(p)_{p\in P}, B(t)_{t\in T}, \left(w^{-/+}{}_{t,p}\right)_{(t,p)\in T\times P} \right)$ with the following properties: All places $p \in P$ are unbranched and have the same colourset $C(p) = \{ high, low \}$ with two token colours. There are exactly two kinds of transitions:

- An AND-transition $t \in T$ has a set of firing modes $B(t) = \{ high, low \}$ with two elements: The high-mode (resp. low-mode) is activated, iff all preplaces of the transition are marked with at least one high-token (resp. low-token), its firing generates one high-token (resp. low-token) on every postplace.
- An XOR-transition $t \in T$ with $n$ preplaces and $m$ postplaces has a set of firing modes $B(t) = \{ b_{(i,j)} \}$ with $n \cdot m$ high-modes and one low-mode: The high-mode with index $(i, j)$, $1 \le i \le n, 1 \le j \le m$ is activated, iff the $i$-th preplace is marked with at least one high-token and all other preplaces with at least one low-token. Firing the high-mode generates a high-token at the $j$-th postplace and a low-token at every other postplace. The low-mode is activated, iff all preplaces are marked with at least one low-token. Firing the low-mode generates a low-token at every postplace.

ii) A *bipolar synchronization system* (BP-system) is a Petri net $BPS = (BPG, \mu)$ with a BP-graph $BPG$ and an initial marking $\mu$ carrying at least one high-token.

Every BP-system decomposes in a canonical way into a subsystem and a quotient. The subsystem is a T-system. The quotient is a restricted free-choice system. Canonical morphisms connect the BP-system with its derived ordinary Petri nets. Therefore we have first to introduce the concept of a morphism between coloured Petri nets. In general a morphism is a map, which respects the structure of the objects in question. What are the relevant structures in case of coloured nets?

The fundamental characteristics of a coloured net are its duality of places versus transitions and token colours versus firing modes. The incidence-functions connect the net components of one type with those of the other type. We will see, that they introduce a further duality between flows and marking classes.

The first duality of places versus transitions originates as a topological structure. The duality is due to the undirected bipartite graph underlying a coloured net. Later on the algebraic structure of a coloured net builds on the topological structure. It introduces the colours and their dualities.

## 2.3 Definition *(Bipartite graph and topology)*

For an undirected bipartite graph $G = (X, ad)$ the adjacency relation $ad \subset X \times X$ defines two topologies on the set $X = P \dot{\cup} T$ of places $P := domain\ ad$ and transitions $T := codomain\ ad$.

i) Open sets of the *P-topology* of $X$ are the place-bordered subsets, i.e. a subset $U \subset X$ is open, iff for all nodes $t \in U \cap T$ holds: If $p \in P$ and $(p, t) \in ad$, then also $p \in U$.

ii) Open sets of the *T-topology* of $X$ are the transition-bordered subsets. i.e. a subset $A \subset X$ is open, iff for all nodes $p \in A \cap P$ holds: If $t \in T$ and $(p, t) \in ad$, then also $t \in A$.

It is easy to check, that both definitions satisfy the requirements of a topology. They are dual in the following sense: The complement of any open set in the P-topology is open in the T-topology and viceversa. Expressed differently: Closed sets in one topology are open in the other. Hence we will focus on only one topology, the P-topology. Here we will consider the duality between open subsets, which are the place-bordered subsets, and closed subsets, which are the transition-bordered subsets.

We now formalize in a second step the algebraic structure of a coloured net. For closed $A \subset X$ and open $U \subset X$ we set

$$\mathcal{B}(A) := \coprod_{t \in A} B(t)_Z \text{ and } \mathcal{C}(U) := \prod_{p \in U} C(p)_Z \text{ ,}$$

which are resp. the coproduct and the product of the free Abelian groups with base resp. $B(t)$ and $C(p)$. The inclusion $\boldsymbol{N} \subset \boldsymbol{Z}$ marks in both groups the cone of non-negative elements. The canonical injections into the coproduct and the canonical projections from the product induce group homomorphisms

$$e_{A_2,A_1} : \mathcal{B}(A_1) \longrightarrow \mathcal{B}(A_2) \text{ and } r_{U_2,U_1} : \mathcal{C}(U_2) \longrightarrow \mathcal{C}(U_2)$$

for each pair of inclusions $A_1 \subset A_2$ and $U_1 \subset U_2$. The two incidence-functions induce two group homomorphisms

$$w^{-/+}{}_{U,A} : \mathcal{B}(A) \longrightarrow \mathcal{C}(U),\ A \subset X \text{ closed and } U \subset X \text{ open,}$$

such that the following two diagrams commute – one for $w^-$, the other for $w^+$:

$$\begin{array}{ccc} \mathcal{B}(A_2) & \xrightarrow{w^{-/+}{}_{U_2,A_2}} & \mathcal{C}(U_2) \\ \uparrow e_{A_2,A_1} & & \downarrow r_{U_2,U_1} \\ \mathcal{B}(A_1) & \xrightarrow{w^{-/+}{}_{U_1,A_1}} & \mathcal{C}(U_1) \end{array}$$

We set $w := w^+ - w^-$, i.e. for all open $U \subset X$ and closed $A \subset X$ the incidence map is a group homomorphism

$$w_{U,A} := w^+{}_{U,A} - w^-{}_{U,A} : \mathcal{B}(A) \longrightarrow \mathcal{C}(U).$$

From now on we represent a coloured net as a tuple $CN = \left(X, \mathcal{C}, \mathcal{B}, w^{-/+}\right)$ and consider the set of nodes $X = P \stackrel{\bullet}{\cup} T$ as a topological space with respect to the P-topology.

## 2.4 Definition (*Flows and marking classes*)

Consider a coloured net $CN = \left(X, \mathcal{C}, \mathcal{B}, w^{-/+}\right)$.

i) For a closed subset $A \subset X$ we define the *flows* of $CN$ on $A$ as the subgroup

$$\mathcal{F}(A) := \ker \left[ w_{U,A} : \mathcal{B}(A) \longrightarrow \mathcal{C}(U) \right]$$

with $U \subset X$ the open kernel of $A$.

ii) For an open subset $U \subset X$ we define the *marking classes* of $CN$ on $U$ as the quotient group

$$\mathcal{M}(U) := coker \left[ w_{U,A} : \mathcal{B}(A) \longrightarrow \mathcal{C}(U) \right]$$

with $A \subset X$ the closed kernel of $U$.

iii) For closed $A_1 \subset A_2$ and open $U_1 \subset U_2$ we have resp. a canonical *restriction*

$$r_{A_2,A_1} : \mathcal{F}(A_2) \longrightarrow \mathcal{F}(A_1)$$

and a canonical *extension*

$$e_{U_2,U_1} : \mathcal{M}(U_1) \longrightarrow \mathcal{M}(U_2).$$

Both maps are compatible with respect to composition:

$$r_{A_3,A_2} \circ r_{A_2,A_1} = r_{A_3,A_1} \text{ and } e_{U_3,U_2} \circ e_{U_2,U_1} = e_{U_3,U_1}$$

With respect to the P-topology arbitrary unions of closed sets are closed. Hence the closed kernel of a set is well-defined. Flows on a closed set are those steps, which do not change the marking in the interior of the set when firing. Marking classes on an open set identify two markings, when one results from the other by firing a step in the interior of the set. Flows and marking classes are well established in the global case $A = U = X$, named resp. T-flows and P-flows of the coloured net. More precisely: P-flows, being linear functionals, are the dual concept of marking classes. Flows and marking classes localize the corresponding global concepts. Specifically, the flows of a single transition are its firing elements, the marking classes of a single place are its token elements.
Also the global concepts of step and marking can be expressed using flows and marking classes.

## 2.5 Remark (*Step and marking*)

Consider a coloured net $CN = \left( X, \mathcal{C}, \mathcal{B}, w^{-/+} \right)$ with $X = P \dot{\cup} T$. Because

$$\mathcal{B}(X) = \mathcal{B}(T) = \mathcal{F}(T) \text{ and } \mathcal{C}(X) = \mathcal{C}(P) = \mathcal{M}(P)$$

a *step* of $CN$ is a non-negative element of $\mathcal{F}(T)$ and a *marking* of $CN$ is a non-negative element of $\mathcal{M}(P)$.

Definition 2.6 selects flows and marking classes as the relevant structures, which a morphism of coloured nets has to conserve. This idea is due to Charles Lakos [Lak1997].

## 2.6 Definition (*Morphism of coloured nets*)

A *morphism* between two coloured nets $CN_X = \left( X, \mathcal{C}_X, \mathcal{B}_X, w_X{}^{-/+} \right)$ and $CN_Y = \left( Y, \mathcal{C}_Y, \mathcal{B}_Y, w_Y{}^{-/+} \right)$ is a tuple

$$(f, f_F, f_M): CN_X \longrightarrow CN_Y$$

formed by

- a continous map $f : X \longrightarrow Y$,
- a family of group homomorphisms between flows

$$f_{F,A} : \mathcal{F}_X\left(f^{-1}(A)\right) \longrightarrow \mathcal{F}_Y(A),\ A \subset Y \text{ closed,}$$

- and a family of group homomorphisms between marking classes

$$f_{M,U} : \mathcal{M}_X\left(f^{-1}(U)\right) \longrightarrow \mathcal{M}_Y(U),\ U \subset Y \text{ open.}$$

These maps satisfy the following conditions:

i) $f_{F,A}$ respects the cone of non-negative elements

ii) $f_{F,A}$ is compatible with respect to restrictions, i.e.

$$r_{Y;A_2,A_1} \circ f_{F,A_2} = f_{F,A_1} \circ r_{X;f^{-1}(A_2),f^{-1}(A_1)},\ A_1 \subset A_2 \subset Y,$$

iii) $f_{M,U}$ is compatible with respect to extensions, i.e.

$$e_{Y;U_2,U_1} \circ f_{M,U_1} = f_{M,U_2} \circ e_{X;f^{-1}(U_2),f^{-1}(U_1)},\ U_1 \subset U_2 \subset Y,$$

iv) The following two diagrams - for $w^-$ and $w^+$ - commute:

$$\begin{array}{ccccccc}
\mathcal{F}_X\left(f^{-1}(A)\right) & \longrightarrow & \mathcal{B}_X\left(f^{-1}(A)\right) & \xrightarrow{w^{-/+}_{X;f^{-1}(U),f^{-1}(A)}} & \mathcal{C}_X\left(f^{-1}(U)\right) & \longrightarrow & \mathcal{M}_X\left(f^{-1}(U)\right) \\
f_{F,A} \downarrow & & & & & & f_{M,U} \downarrow \\
\mathcal{F}_Y(A) & \longrightarrow & \mathcal{B}_Y(A) & \xrightarrow{w^{-/+}_{Y;U,A}} & \mathcal{C}_Y(U) & \longrightarrow & \mathcal{M}_Y(U)
\end{array}$$

for all $A \subset Y$ and open $U \subset Y$ and with canonical homorphisms between the subgroups and the quotient groups in each row.

A morphism is *discrete*, iff its component $f : X \longrightarrow Y$ is a discrete map, i.e. iff the induced topology on each fibre is the discrete topology.

The map $f : X \longrightarrow Y$ is discrete, iff the inverse image of a place contains only places and the inverse image of a transition contains only transitions. All important morphisms of the present paper are discrete, in particular all morphisms from Definition 2.9. A fibre of a general map $f : X \longrightarrow Y$ is not discrete, iff it is a non-empty subnet, containing places as well as transitions.

For a discrete morphism every firing element and every colour element within a fibre is also resp. a flow and a marking class of the fibre. A morphism of coloured nets maps flows and marking classes. Hence there is no difficulty to map arbitrary steps, markings and occurrence sequences along a discrete morphism.
For a non-discrete morphism

$$(f, f_F, f_M): CN_X \longrightarrow CN_Y$$

one has to pay some attention. Set $X = P_X \dot{\cup} T_X$ and $Y = P_Y \dot{\cup} T_Y$. All subsets of $T_X$ are closed and all subsets of $P_X$ are open. We have

$$\mathcal{B}_X(X) = \mathcal{B}_X(T_X) = \mathcal{B}_X(T_X - f^{-1}(Y)) \oplus \mathcal{B}_X(T_X \cap f^{-1}(Y)) =$$

$$\mathcal{B}_X(T_X - f^{-1}(Y)) \oplus \mathcal{B}_X(T_X \cap f^{-1}(P_Y)) \oplus \mathcal{B}_X(T_X \cap f^{-1}(T_Y))$$

A step $\tau \in \mathcal{B}_X(X)$ is called *saturated with respect to* $(f, f_F, f_M)$, if it decomposes as

$$\tau = \tau_1 + \tau_2 + \tau_3 \in \mathcal{B}_X(T_X - f^{-1}(Y)) \oplus \mathcal{B}_X(T_X \cap f^{-1}(P_Y)) \oplus \mathcal{B}_X(T_X \cap f^{-1}(T_Y))$$

with the third summand being a flow, i.e. with

$$\tau_3 \in \mathcal{F}_X(f^{-1}(T_Y)) \subset \mathcal{B}_X(f^{-1}(T_Y)) = \mathcal{B}_X(T_X \cap f^{-1}(T_Y)).$$

We extend $f_F$ to saturated steps $\tau \in \mathcal{B}_X(X)$ by defining

$$f_F(\tau) := f_F(\tau_3) \in \mathcal{F}_Y(T_Y) = \mathcal{B}_Y(Y).$$

An occurrence sequence of $CN_X$ is *saturated with respect to* $(f, f_F, f_M)$, if it is a catenation of saturated steps. Due to its non-negativity the map $f_F$ extends to a map of saturated occurrence sequences.
Similarly on the level of markings: The subset $f^{-1}(P_Y)$ of $X$ is open and defines the restriction

$$r : \mathcal{C}_X(X) \longrightarrow \mathcal{C}_X(f^{-1}(P_Y)).$$

Using the quotient map

$$\mathcal{C}_X(f^{-1}(P_Y)) \longrightarrow \mathcal{M}_X(f^{-1}(P_Y)), \mu \mapsto [\mu],$$

we extend $f_M$ from a map of marking classes to a map of global markings

$$f_M : \mathcal{C}_X(X) \longrightarrow \mathcal{C}_Y(Y) = \mathcal{M}_Y(P_Y), \mu \mapsto f_M([r(\mu)]).$$

## 2.7 Definition *(Morphism of coloured Petri nets)*

A *morphism of coloured Petri nets*

$$(f, f_F, f_M): (CN_X, \mu_X) \longrightarrow (CN_Y, \mu_Y)$$

is a morphism $(f, f_F, f_M): CN_X \longrightarrow CN_Y$ of the underlying coloured nets with $f_M(\mu_X) = \mu_Y$.

## 2.8 Remark *(Morphism of coloured Petri nets)*

i) With respect to a discrete morphism every occurrence sequence is saturated.

ii) Consider a general morphism between Petri nets

$$(f, f_F, f_M): PN_X \longrightarrow PN_Y.$$

If a saturated occurrence sequence $\sigma_X$ of $PN_X$ fires according to $\mu_{X,1} \xrightarrow{\sigma_X} \mu_{X,2}$, then its image $\sigma_Y := f_F(\sigma_X)$ fires according to $f_M(\mu_{X,1}) \xrightarrow{\sigma_Y} f_M(\mu_{X,2})$.

A bipolar synchronization system is a coloured Petri net. It is linked with a series of ordinary Petri nets. Each link is given by a discrete morphism.

## 2.9 Definition *(Petri nets derived from BP-systems)*

Consider a BP-system $BPS = (N, \mu)$.

i) The *flat-system* of $BPS$ is the ordinary Petri net $BPS^{flat} = (N^{flat}, \mu^{flat})$ together with the morphism

$$BPS^{flat} \xrightarrow{col} BPS,$$

which are defined as follows: Places of $N^{flat}$ are the token elements of $N$, transitions of $N^{flat}$ are the firing elements of $N$. Places which unfold high-token elements are called high-places. Analogously we define low-places, high-transitions and low-transitions. A place $(p, c) \in \mathcal{C}(p)$ and a transition $(t, b) \in \mathcal{B}(t)$ are joined in $N^{flat}$ by a directed arc, iff there exists an arc in $N$ with the same orientation between $p$ and $t$. The map $col$ operates as a projection on places and transitions $col(p, c) := p$ for $c \in \mathcal{C}(p)$ and $col(t, b) := t$ for $b \in \mathcal{B}(t)$ and as the identity on token elements and firing elements. The initial marking is defined as $\mu^{flat} := col^{-1}(\mu)$.

ii) The *low-system* of $BPS$ is the coloured Petri net $BPS^{low} = (N^{low}, \mu^{low})$ together with the morphism

$$BPS^{low} \xrightarrow{low} BPS,$$

which are defined as follows: $N^{low}$ is the subnet of $N$ generated by all *low*-token elements and all *low*-firing elements: $N^{low}$ has the same places and transitions as $N$, but all token elements and firing elements of $N^{low}$ have only a single colour $\{low\}$. The initial marking $\mu^{low}$ of $N^{low}$ is made up of the low-tokens of $\mu$. The net inclusion $N^{low} \xrightarrow{\subset} N$ induces the morphism $BPS^{low} \xrightarrow{low} BPS$.

iii) The *high-system* of $BPS$ is the cokernel $BPS^{high} = \left(N^{high}, \mu^{high}\right)$ of the mapping from part ii), i.e. the quotient $BPS^{high} := BPS/low\left(BPS^{low}\right)$, together with the canonical quotient map

$$BPS \xrightarrow{high} BPS^{high} :$$

The net $N^{high}$ has the same places and transitions as $N$, but token elements of $N^{high}$ are restricted to high-token elements of $N$ and firing elements are restricted to high-modes of $N$. The incidence-maps of $N^{high}$ are induced as quotients by the incidence-maps of $N$.

iv) The *uncolouring* $N \xrightarrow{uncol} N^{low}$ is a net morphism, operating as the identity on places and transitions and mapping resp. token and firing elements as $uncol(t,b) := (t, low)$ and $uncol(p,c) := (p, low)$.

The *skeleton* of $BPS$ is the Petri net $BPS^{skel} := \left(N^{low}, \mu^{skel}\right)$, $\mu^{skel} := uncol(\mu)$, together with the morphism

$$BPS \xrightarrow{skel} BPS^{skel},$$

induced by the uncolouring.

All morphisms from Definition 2.9 are discrete. The low-system $BPS^{low}$ is a subsystem of $BPS$, but it is even a retraction, because the low-morphism has a section $BPS^{low} \xrightarrow{uncol \circ low = id} BPS^{low}$. The high-system has been defined as the cokernel of a Petri net morphism. Note that the quotient is not defined on the level of token colours and firing modes, i.e. in the category of sets, but in the category of Abelian groups. Any occurrence sequence $\sigma$ of $BPS$ induces an occurrence sequence $high(\sigma)$ of $BPS^{high}$ by skipping all low-modes.

On the flattening of the low-system $BPS^{flat,low} := col^{-1}\left(BPS^{low}\right) \subset BPS^{flat}$ the morphism $col$ induces an isomorphism

$$col : BPS^{flat,low} \xrightarrow{\cong} BPS^{low}.$$

Hence we will consider the low-system as an ordinary Petri net. Also the quotient

$$BPS^{flat,high} := coker\left[BPS^{flat,low} \xrightarrow{\subset} BPS^{flat}\right],$$

the *flat high-system*, is an ordinary Petri net. The colouring induces a morphism $BPS^{flat,high} \longrightarrow BPS^{high}$, which maps bijectively activated occurrence sequences of both Petri nets. Hence one of the two Petri nets is safe and live, iff also the other has these properties. For all objectives of the present paper we may consider the restricted free choice-system $BPS^{flat,high}$ instead of $BPS^{high}$. The following diagram of Petri net morphisms commutes

$$\begin{array}{ccccc} BPS^{flat,low} & \xrightarrow{low} & BPS^{flat} & \xrightarrow{high} & BPS^{flat,high} \\ \cong\downarrow col & & \downarrow col & & \downarrow col \\ BPS^{low} & \xrightarrow{low} & BPS & \xrightarrow{high} & BPS^{high} \end{array}$$

**Fig. 1**. BP-system and its derived Petri nets

Setting $BPS = (N, \mu)$ and $BPS^{flat,high} = \left(N^{flat,high}, \mu^{flat,high}\right)$ the morphisms mentioned above map bijectively the places of $N$ to the places of $N^{flat,high}$: We denote the image of a place $p \in N$ by

$$p^{high} := col^{-1}(\ high(\ p\ )\ ) \in N^{flat,high}\ ^{1}.$$

Similarly they map injectively all AND-transitions into the transitions of $N^{flat,high}$: We denote the image of an AND-transition $t_{AND} \in N$ by

$$t_{AND}{}^{high} := col^{-1}(\ high\ (\ t_{AND}\ )\ ) \in N^{flat,high}\ .$$

The inverse image

$$col^{-1}(\ high(\ t_{XOR}\ )\ ) \in N^{flat,high}$$

of an XOR-transition $t_{XOR} \in N$ is a set of transitions $t_{XOR}{}^{high} \in N^{flat,high}$, namely the $n \cdot m$ high-modes of $t_{XOR}$. This correspondence on the level of places and transitions induces a bijective map between paths $\alpha \subset N$ and paths $\alpha^{high} \subset N^{flat,high}$ with prescribed initial and final node.

## 2.10 Remark *(Petri nets as sheaves and cosheaves)*

The technical details referring to flows, marking classes and maps of a coloured net $CN = \left(X, \mathcal{C}, \mathcal{B}, w^{-/+}\right)$ can be expressed in a more compact form, cf. [Weh2006]: The two families $\mathcal{C}$ and $\mathcal{B}$ form resp. a sheaf on $X$ with respect to the P-topology and a cosheaf with respect to the T-topology. Flows form a sheaf $\mathcal{F}$ with respect to the T-topology, while marking

---

[1] We abbreviate resp. $p^{flat,high}, t^{flat,high} \in N^{flat,high}$ by $p^{high}, t^{high} \in N^{flat,high}$

classes form a cosheaf $\mathcal{M}$ with respect to the P-topology. For a morphism between coloured nets one first needs a continous map $f : X \longrightarrow Y$ of the underlying topological spaces. After introducing the direct images $f_*(\mathcal{F}_X)$ and $f_*(\mathcal{M}_X)$ one secondly postulates two morphisms between resp. sheaves and cosheaves

$$f_T : f_*(\mathcal{F}_X) \longrightarrow \mathcal{F}_Y \text{ and } f_M : f_*(\mathcal{M}_X) \longrightarrow \mathcal{M}_Y,$$

which are compatible with the incidence-maps from cosheaves to sheaves.

# 3 Liveness of BP-systems: Necessary condition

In this chapter we consider a safe and live BP-system. We derive necessary properties of its high-system and skeleton. The result and its proof are well known and mostly contained in [GT1984]. Our method of proof uses the concept of Petri net morphisms to investigate the properties of the Petri nets derived from the BP-system. This method prepares the proof of the converse result in Chapter 4.

## 3.1 Lemma *(Safe BP-system)*

A BP-system is safe, if its skeleton is safe.

**Proof.** Because the morphism $BPS \xrightarrow{skel} BPS^{skel}$ maps activated occurrence sequences, it maps any reachable marking of $BPS$ to a reachable marking of $BPS^{skel}$. If no reachable marking of $BPS^{skel}$ marks a place with more than a single token, the same holds true for $BPS$, q. e. d.

The lifting problem considers the converse situation: Under which assumptions does a Petri net morphism $PN_1 \xrightarrow{f} PN_2$ have the *lifting property*, i.e. given an activated occurrence sequence $\sigma_2$ of $PN_2$, when does there exist an activated occurrence sequence $\sigma_1$ of $PN_1$ with $f(\sigma_1) = \sigma_2$ ? For the skeleton we will solve the lifting problem with Lemma 3.3, for the high-system with Corollary 3.5.

## 3.2 Definition *(Live, dead, deadlock)*

Consider a BP-system $BPS = (N, \mu)$.

i) A firing element of $N$ is *live at the marking* $\mu$, iff for every reachable marking $\mu_1$ the BP-system $(N, \mu_1)$ has a reachable marking, which activates the given firing element. $BPS$ is *live with respect to all high-modes*, iff every high-firing element is live at $\mu$.

ii) A transition of $N$ is *high-live at the marking* $\mu$, iff it has a high-firing element, which is live at $\mu$. $BPS$ is high-live, if each transition is high-live at $\mu$.

iii) A transition of $N$ is *dead at a reachable marking* $\mu_1$, if no reachable marking of the BP-system $(N, \mu_1)$ activates any firing mode of the given transition. A reachable marking $\mu_1$ is *dead*, if all transitions of $BPS$ are dead at $\mu_1$. $BPS$ is *dead*, if the initial marking $\mu$ is dead.

iv) A reachable marking $\mu_1$ of $BPS$ has a *deadlock-transition* $t \in N$, if either $t$ is an AND-transition with at least one high-marked and one low-marked preplace at $\mu_1$ or if $t$ is a XOR-transition with at least two high-marked preplaces at $\mu_1$. $BPS$ is *deadlock-free*, if no reachable marking has a deadlock-transition.

## 3.3 Lemma *(Lifting property of the skeleton)*

For a deadlock-free BP-system $BPS$ the skeleton $BPS \xrightarrow{skel} BPS^{skel}$ has the lifting property.

More precisely: Let $\gamma$ be a path in $BPS$, which starts with a place high-marked at a reachable marking $\mu$, and let $skel(\mu) \xrightarrow{\sigma^{skel}} \mu_1^{skel}$ with $\gamma^{skel} \subset supp(\sigma^{skel})$ be an occurrence sequence of $BPS^{skel}$. Then $\sigma^{skel}$ has a lift $\sigma$ to $BPS$, such that every transition from $\gamma$ fires a high-mode.

**Proof.** Set $BPS = (N, \mu)$ and consider an activated occurrence sequence $\sigma^{skel}$ of $BPS^{skel}$. We may assume, that $\sigma^{skel}$ is a single transition $t^{skel} \in N^{skel}$ firing according to $skel(\mu) \xrightarrow{\sigma^{skel}} \mu_1^{skel}$. All preplaces of the corresponding transition $t \in N$ are marked. Due to the deadlock-freeness of $BPS$ the marking $\mu$ activates a firing mode $b \in B(t)$ of $BPS$ with $\sigma^{skel} = skel(t, b)$. In case of an XOR-transition $t$ the firing mode can be choosen according to the demand of $\gamma$. Hence the occurrence sequence $\sigma := (t, b)$ of $BPS$ is a lift of $\sigma^{skel}$, q. e. d.

## 3.4 Corollary *(Live or dead with deadlock)*

For a BP-system $BPS$ with a safe skeleton holds the equivalence:

1. $BPS$ is high-live
2. No reachable marking of $BPS$ is dead
3. $BPS$ is deadlock-free and the skeleton is live.

**Proof.** $1 \Rightarrow 2$ The proof is obvious, because liveness always implies non-deadness.

$2 \Rightarrow 3$ Due to the assumption no reachable marking of $BPS = (N, \mu)$ has a deadlock-transition. In order to verify the liveness of the T-system $BPS^{skel}$, it suffices to show, that every reachable marking $\mu_1^{skel}$ of $BPS^{skel}$ activates at least one transition $t^{skel}$. By Lemma 3.3 every occurrence sequence $\sigma^{skel}$ of $BPS^{skel}$, firing according to $\mu^{skel} \xrightarrow{\sigma^{skel}} \mu_1^{skel}$, lifts to an occurrence sequence $\mu \xrightarrow{\sigma} \mu_1$ of $BPS$ with $skel(\mu_1) = \mu_1^{skel}$. By assumption $\mu_1$ activates at least one firing mode $(t, b) \in N$, hence $\mu_1^{skel}$ activates the transition $t^{skel} := skel(t, b)$.

$3 \Rightarrow 1$ In the proof we will use the safeness of the skeleton. Consider a marking $\mu_1$, activated in $BPS$ due to the firing of an occurrence sequence $\mu \xrightarrow{\sigma} \mu_1$, and a transition $t \in N$. Because the initial marking $\mu$ contains at least one high-token, the same holds true for $\mu_1$ and there exists a transitions $t_1 \in N$ with a preplace high-marked at $\mu_1$. By assumption the transition $skel(t_1) \in N^{skel}$ can be activated by a minimal occurrence sequence $skel(\mu_1) \xrightarrow{\sigma_1^{skel}} \mu_2^{skel}$, such that $skel(t_1)$ is the only transition of $N^{skel}$ activated at $\mu_2^{skel}$. By Lemma 3.3 the occurrence sequence $\sigma_1^{skel}$ lifts to $\mu_1 \xrightarrow{\sigma_1} \mu_2$ and $\mu_2$ activates a high-mode of $t_1$. The safe and live T-system $(N^{skel}, \mu_2^{skel})$ contains an unmarked path $\gamma^{skel}$ from $skel(t_1)$ to $skel(t)$ as well as a minimal activated occurrence sequence $\mu_2^{skel} \xrightarrow{\sigma_2^{skel}} \mu_3^{skel}$, $\gamma^{skel} \subset supp(\sigma_2^{skel})$, which activates $skel(t)$. By Lemma 3.3 it lifts to $\mu_2 \xrightarrow{\sigma_2} \mu_3$ and $\mu_3$ activates a high-mode of $t$, q. e. d.

**Note.** The essential step „$3 \Rightarrow 1$" in the proof of Corollary 3.4 has already been demonstrated by Genrich and Thiagarajan, also Corollary 3.5 is due to them ([GT1984], Theor. 2.12, Lemma 3.10). In the following the concatenation $\sigma_1 \cdot \sigma_2$ of two occurrence sequences means to execute first $\sigma_1$ on the left and afterwards $\sigma_2$. While composition $\alpha_2 \circ \alpha_1$ of two paths means to traverse first $\alpha_1$ on the right and afterwards $\alpha_2$. Here we have used the common notation for the composition of maps, because a path is map.

## 3.5 Corollary *(Lifting property of the high-system)*

For a deadlock-free BP-System $BPS$ with safe and live skeleton the high-system $BPS \xrightarrow{high} BPS^{high}$ has the lifting property.

**Proof.** Set $BPS = (N, \mu)$ and denote by $BPS^{flat} = (N^{flat}, \mu^{flat})$ the flat high-system. It suffices to prove the analogous statement for the morphism $BPS^{flat} \xrightarrow{high} BPS^{flat,high}$ with cokernel

$$BPS^{flat,high} = coker\left[BPS^{flat,low} \xrightarrow{low} BPS^{flat}\right].$$

In $BPS^{flat,high} = (N^{flat,high}, \mu^{flat,high})$ we consider an activated occurrence sequence $\sigma^{high}$ firing according to $\mu^{flat,high} \xrightarrow{\sigma^{high}} \mu_1^{flat,high}$. Without loss of generality $\sigma^{high}$ is a single transition, i.e. $\sigma^{high} = high(\ \sigma_h\ )$ with $\sigma_h := (\ t, b\ ) \in N^{flat}$ with a transition $t \in N^{flat}$ and a high-mode $b \in B(t)$. For the proof we shall catenate $\sigma_h$ with a second occurrence sequence $\sigma_l$ of $BPS^{flat}$, such that $\sigma := \sigma_l \cdot \sigma_h$ is activated in $BPS^{flat}$ and still satisfies $\sigma^{high} = high(\sigma)$. Hence the image $high\ (\sigma_l)$ is the empty sequence and we have to find $\sigma_l$ as a suitable occurrence sequence of the low-subsystem $BPS^{flat,low} \subset BPS^{flat}$. In case $\sigma_h$ is activated at $\mu^{flat}$, we can choose $\sigma_l$ as the empty sequence.

Otherwise $high\ (\ \sigma_h\ )$ is activated at $\mu^{high}$, but $\sigma_h$ lacks activation at $\mu$.

Therefore $\sigma_h = (\ t_{XOR}, b\ )$ with a closing XOR-transition $t_{XOR}$, a high-mode $b \in B(t_{XOR})$ and a high-marked preplace $p \in pre(t_{XOR}) \subset N$. Lemma 3 implies that $BPS$ is safe. Hence no preplace of $t_{XOR}$ is marked with more than one token. Moreover no preplace of $t_{XOR}$ different from $p$ is high-marked due to the deadlock-freeness of $BPS$. Eventually, due to the lacking activation of $(\ t_{XOR}, b\ )$ the transition $t_{XOR}$ has at least one unmarked preplace. In order to activate $(\ t_{XOR}, b\ )$ at a reachable marking, it is necessary to mark any of the unmarked preplaces of $t_{XOR}$ with a low-token. The skeleton $BPS^{skel} = (N^{skel}, \mu^{skel}\ )$ is live. Hence there exists a minimal occurrence sequence $\mu^{skel} \xrightarrow{\sigma^{skel}} \mu_0^{skel}$ of $BPS^{skel}$ with $\mu_0^{skel}$ activating the transition $t^{skel} := skel(\ t_{XOR}, b\ ) \in N^{skel}$. It lifts to an occurrence sequence $\mu \xrightarrow{\sigma^{low}} \mu_0$ of $BPS$ with $\mu_0$ activating a firing element $(\ t_{XOR}, b'\ )$, $b' \in B(t_{XOR})$, cf. Lemma 3.3. Because the minimal occurrence sequence $\sigma^{skel}$ does not contain the transition $t^{skel}$, the firing element $(\ t_{XOR}, b\ )$ does not belong to $\sigma^{low}$. Hence its preplace $p$ remains high-marked at $\mu_0$. Due to the deadlock-freeness of $BPS$ all other preplaces of $t_{XOR}$ must be low-marked. We obtain $(\ t_{XOR}, b'\ ) = (\ t_{XOR}, b\ )$.

Claim: Every firing mode of $\sigma^{low}$ is a low-mode, i.e. $\sigma^{low}$ belongs to the low-system $BPS^{low}$. For the proof note, that every transition from $\sigma^{skel}$ with an elementary token-free path to $t^{skel}$ fires exactly once. Hence $\sigma^{low}$ contains only firing modes of transitions with a token-free path to $t_{XOR}$. Moreover, all firing elements of $\sigma^{low}$ belong to pairwise different transitions. Under the assumption, that $\sigma^{low}$ contains the high-mode of a transition, we select an elementary path $\gamma \subset supp(\sigma^{low})$ from a high-marked preplace of that transition to $t_{XOR}$. According to Lemma 3.3 we can choose the lift $\sigma^{low}$, such that its firing creates a high-token on the preplace $p_n \neq p$ of $t_{XOR}$. Hence $t_{XOR}$ is a deadlock-transition at $\mu_0$. This contradicts the deadlock-freeness of $BPS$ and proves, that every firing element of $\sigma^{low}$ is a low-mode. We set $\sigma_l := uncol(\sigma^{low})$. The catenation $\sigma := \sigma_l \cdot \sigma_h$ is an activated occurrence sequence of $BPS^{flat}$ and lifts $\sigma^{high}$, because

$$high(\sigma) = high(\sigma_l) \cdot high(\sigma_h) = high(\sigma_h) = \sigma^{high} \text{, q. e. d.}$$

The following Theorem 3.6 is essentially due to Genrich and Thiagarajan ([GT1984], Theor. 3.13) and is one of their main results.

## 3.6 Theorem *(Liveness and safeness of BP-systems)*

For a high-safe and live BP-system the skeleton is safe and live and the high-system is safe and live without frozen tokens.

**Proof.** i) Denote by $BPS = (N, \mu)$ the given BP-system. Safeness of $BPS^{skel}$ follows from Lemma 3.3, and liveness of $BPS^{skel}$ follows from Corollary 3.4. Safeness of $BPS^{high}$ follows from Corollary 3.5. Because the high-system is a safe strongly connected free-choice system, its deadlock-freeness is equivalent to liveness ([DE1995], Theor. 4.31). For an indirect proof of the deadlock-freeness we assume, that $BPS^{high} = (N^{high}, \mu^{high})$ has a reachable dead marking $\mu_1^{high}$. It is generated by an occurrence sequence $\mu^{high} \xrightarrow{\sigma^{high}} \mu_1^{high}$, which lifts to $\mu \xrightarrow{\sigma} \mu_1$ by Corollary 3.5. Because $BPS$ is high-live by assumption, the marking $\mu_1$ activates a high-firing element $(t, b)$ of at least one transition $t \in N$. Its image $high(t, b) \in N^{high}$ is a transition of the high-system, which is activated at $\mu_1^{high}$, a contradiction.

ii) Exclusion of frozen tokens: For an indirect proof we assume the existence of a reachable marking $\mu_1^{high}$ and a place $p^{high} = high(p) \in N^{high}$ marked at $\mu_1^{high}$ with a frozen token. Denote by $\sigma^{high}$ an infinite activated occurrence sequence of $(N^{high}, \mu_1^{high})$, which does not move the frozen token. By Corollary 3.5 it lifts to an infinite activated occurrence sequence $\sigma$ of the BP-system $(N, \mu_1)$, which does not move the token at the place $p \in N$. Now $skel(\sigma)$ is an infinite activated occurrence sequence of the skeleton $BPS^{scel}$ with a frozen token at the place $skel(p) \in N^{skel}$. But the skeleton is a safe and live T-system as already proved in part i). Hence it has no frozen tokens, cf. Lemma 1.5. This contradiction shows, that also the high-system has no frozen tokens, q. e. d.

## 3.7 Corollary *(Liveness with respect to all high-modes)*

A safe and high-live BP-system is live with respect to all high-modes.

**Proof.** By Corollary 3.4 high-liveness of a BP-system implies its deadlock-freeness. By Theorem 3.6 and Corollary 3.5 every activated occurrence sequence of the high-system lifts to an activated occurrence sequence of the given BP-system, q. e. d.

# 4 Liveness of BP-systems: Sufficient condition

In the present chapter we prove Theorem 4.6 as the main result of the paper. It is the converse of Theorem 3.6. Because liveness of a BP-system follows from its deadlock-freeness, it suffices to focus on deadlock-freeness. Our proof will be indirect. Therefore we first study dead BP-systems. Without loss of generality we concentrate on BP-systems with *binary* transitions. A binary transition either has a single preplace and two postplaces (*opening* transition) or two preplaces but a single postplace (*closing* transition). One can replace an arbitrary BP-system by a BP-system with only binary transitions without changing lifeness and safeness. This substitution can be formalized by Petri net morphisms: One uses transition refinements, which replace a given transition with an arbitrary number of pre- or postplaces by a T-subnet with binary transitions. Because the fibers of the morphism are no longer discrete, one now has to consider the general definition of Petri net morphisms. Not stating the contrary from now on we assume all transitions as binary transitions.

## 4.1 Lemma *(Dead BP-system)*

For a dead BP-system $BPS$ the following holds:

i) All preplaces of an opening transition are unmarked. No closing transition of the flat high-system is activated.

ii) If the high-system is safe and live, then $BPS$ contains at least one closing XOR-transition with one high-marked and one unmarked preplace and $BPS$ contains no closing XOR-transition with two marked preplaces. In particular, $BPS$ contains no closing XOR-transition in deadlock.

iii) If the skeleton and the high-system are safe and live, then the only transitions activated in the skeleton have the form $skel(t_{AND})$ with a closing AND-transition $t_{AND}$ in deadlock (cf. Fig. 2). Hence $BPS$ contains at least one closing AND-transition in deadlock.

**Proof.** ad i) Any opening transition with a marked preplace would be activated, contradicting the deadness of $BPS = (N, \mu)$. Closing transitions of $N^{flat,high}$ correspond bijectively to closing *AND*-transitions of $N$ . If the former were activated, the latter would be activated, too.

ad ii) If $BPS^{high}$ is safe and live, then at least one transition $high(t) \in BPS^{high}$ must be activated. According to part i) the transition $t \in BPS$ neither is an opening transition nor a closing AND-transition. Hence $t$ is a closing XOR-transition with at least one high-marked preplace. The other preplace is unmarked: A high-token would contradict the safeness of $BPS^{high}$, a low-token would activate $t$, contradicting the deadness of $BPS$ .

ad iii) If $BPS^{skel}$ is safe and live, then at least one transition $skel(t) \in BPS^{skel}$ must be activated. Due to part i) the corresponding transition $t \in BPS$ must be a closing transition, with both preplaces marked and according to part ii) it cannot be an XOR-Transition. Therefore $t$ is an AND-transition, which is not activated. Hence $t$ must be in deadlock, q. e. d.

Our investigation of a dead BP-system is based on the concept of an AND/XOR-chain and on the concept of a deadlock-configuration from Definition 4.2.

## 4.2 Definition *(Deadlock-configuration)*

Consider a BP-system $BPS = ( N, \mu )$.

i) An $AND/XOR$ *-chain of length* $n \geq 1$ of $BPS$ , leading from a closing XOR-transition $t_{XOR}$ to a closing AND-transition $t_{AND}$ , is a family

$$Ch_{AND/XOR} = ( t_k, N_k, p_k )_{k=0,...,n},$$

consisting of the transition $t_0 = t_{XOR}$, closing AND-transitions $t_k$ of $N$ , $k = 1,...,n-1$, and the transition $t_n = t_{AND}$ , together with a family of basic components of the flat high-system $N_k \subset BPS^{flat,high}$ , $k = 0,...,n$: For $k \geq 0$ the distinguished marked place $p_k \in N_k$ satisfies $p_k^{high} \in pre\left( t_k^{high} \right)$ and for $k < n$ holds $t_k^{high} \in N_{k+1}$ .

ii) A *deadlock-configuration of size* $m \geq 1$ of $BPS$ is a family

$$( Ch_{i,AND/XOR}, \beta_i ),\ i = 0, ..., m-1,$$

of AND/XOR-chains $Ch_{i,AND/XOR}$ leading from $t_{i,XOR}$ to a deadlock-transition $t_{i,AND}$ , together with elementary token-free paths $\beta_i$, $i = 0, ..., m-1$, from $t_{i,AND}$ to the postplace of $t_{i+1,XOR}$ .

A deadlock-configuration is *minimal*, if $BPS$ has no deadlock-configuration of smaller size.

Here and in the following any computation with indices from the set $\{ 0, ..., m-1 \}$ has to be understood $modulo\ m$.

For an AND/XOR-chain $Ch_{AND/XOR} = ( t_i, N_i, p_i )_{i=0,...,n}$ exactly one preplace of the transition $t_0 = t_{XOR}$ is high-marked. The flat high-system $BPS^{flat,high}$ has an activated high-mode $t_{XOR}^{high}$ of $t_{XOR}$ . Moreover the activation of each transition $t_{k_0}^{high}$ , $0 < k_0 \leq n$ , presupposes the firing of $t_{XOR}^{high}$ and of all transitions $t_k^{high}$ , $k < k_0$. For a deadlock-configuration $(Ch_{i,AND/XOR}, \beta_i)_{i=0,...,m-1}$ each transition $skel(t_{i,AND})$ but no transition $skel(t_{i,XOR})$ of the skeleton $BPS^{skel}$ is activated. The activation of $skel(t_{i+1,XOR})$ presupposes the firing of $skel(t_{i,AND})$. Hence the two transitions $t_{i,AND}$ and $t_{j,XOR}$ block each other in $BPS$ .

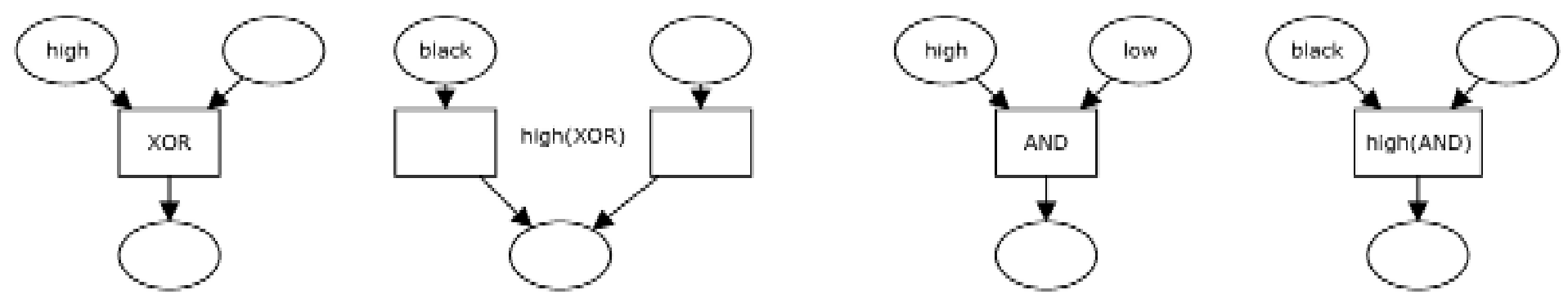

**Fig. 2**. Initial and final transition of an AND/XOR-chain within a deadlock-configuration

## 4.3 Lemma *(AND/XOR-chain in the absence of frozen tokens)*

For an AND/XOR-chain $(t_i, N_i, p_i)_{i=0,\dots,n}$ of a BP-System $BPS = (N, \mu)$ with a safe and live high-system without frozen tokens holds: Each T-component of $N^{flat,high}$ through a given high-mode of a transition $t_{i_0}$, $0 \le i_0 \le n$, contains all transitions $t_i^{high}$, $i_0 < i \le n$.

**Proof.** Denote by $N_T \subset N^{flat,high}$ a T-component passing through the given high-mode $t^{high}$ of $t_{i_0}$. It suffices to consider only the case $i = i_0 + 1$ and to prove $t_i^{high} \in N_T$. We denote by $p_i$ the marked preplace of $t_i^{high}$, it is unbranched in forward direction. For the T-system $(N_T, \mu_T)$, $\mu_T := \mu^{flat,high} \mid N_T$, one of the following two cases holds.

i) $(N_T, \mu_T)$ is live. Then $(N_T, \mu_T)$ has an activated infinite occurrence sequence. It is also an activated occurrence sequence of $BPS^{flat,high}$. All tokens of the initial marking $\mu^{flat,high}$ mark places of $N_T$, otherwise $BPS^{flat,high}$ would have a frozen token. In particular $p_i \in N_T$, which implies $t_i^{high} \in N_T$.

ii) $(N_T, \mu_T)$ is not live. According to Theorem 1.1 there exists a minimal activated occurrence sequence of $BPS^{flat,high}$, which activates $N_T$ without firing any transition from $cl(N_T)$. If $t_i^{high} \notin N_T$, then also $p_i \notin N_T$. In order to activate $N_T$, one has to fire $t_i^{high}$, because the token at $p_i$ would otherwise be frozen. As remarked before, firing $t_i^{high}$ presupposes firing the activated high-mode $t_{i_0}^{high}$ of $t_{i_0}$. But $t_{i_0}^{high} \in N_T$ for $i_0 > 0$, while for $i_0 = 0$ also a transition from

$$post\left(post\left(t_0^{high}\right)\right) \in cl\left(post\left(t_0^{high}\right)\right) \subset cl(N_T)$$

has to fire. In both cases we obtain a contradiction, q. e. d.

The following Algorithm 4.4 considers a dead BP-system with a safe and live high-system. After input of a deadlock-transition $t_{AND}$ it returns an AND/XOR-chain ending at $t_{AND}$. The construction proceeds by backward tracking.

## 4.4 Algorithm *(Construction of an AND/XOR-chain)*

Input: A dead BP-system $BPS$ with a safe and live high-system and a closing AND-transition $t_{AND}$ in deadlock.

Output: AND/XOR-chain $(t_i, N_i, p_i)_{i=0,...,n}$ with final transition $t_n = t_{AND}$.

| | |
|---|---|
| set $j=0$, set termination = false | |
| set $t^{j,fin} = t_{AND}{}^{high} \in N^{flat,high}$ | |
| select a basic component $N^j \subset BPS^{flat,high}$ through the marked preplace $p^j$ of $t^{j,fin}$ | |
| while termination == false | |
| | set $j=j+1$, $t^{j,ini} = t^{j-1,fin} \in N^{flat,high}$ |
| | set $q \in pre\left(t^{j,ini}\right) \in N^{flat,high}$ the unmarked preplace |
| | If $q \in N^k$ for an index $k<j$, set $N^j = N^k$,<br>otherwise select an arbitrary basic component $N^j \subset BPS^{flat,high}$ through $q$ |
| | set $p^j \in N^j$ the distinguished marked place of $N^j$ |
| | set $t^{j,fin} = post\left(p^j\right) \in N^{flat,high}$ |
| | if $t^{j,fin}$ is a high mode of a XOR-transition of $N$, set termination = true |
| set $n=j$, for $i=0,...,n$ set<br>$t_i \in BPS$ with $high\left(t_i\right) = col\left(t^{n-i,fin}\right) \in BPS^{high}$, $N_i = N^{n-i}$, $p_i = p^{n-i}$ | |
| return $(t_i, N_i, p_i)_{i=0,...,n}$ | |

**Proof of correctness.** Set $BPS = (N, \mu)$. The algorithm operates on the flat high-system $BPS^{flat,high} = \left(N^{flat,high}, \mu^{flat,high}\right)$.

i) Well-definedness: For every loop with index $j \geq 1$ after increment at the start of the loop body we assume that the transition $t^{j,ini} = t^{j-1,fin} \in N^{flat,high}$ has exactly one unmarked preplace $q \in pre\left(t^{j,ini}\right)$. This assumption is satisfied for $t^{1,ini} = t_{AND}{}^{high}$ in the first loop with $j=1$. Because the high-system is safe and live, it has a basic component passing through the place $q$. According to Lemma 4.1 every marked place of $N^{flat,high}$ has only one posttransition. Hence $t^{j,fin} := post\left(p^j\right)$ is well-defined. If $t^{j,fin}$ is a high-mode of a XOR-transition $t_{XOR} \in N$, the algorithm terminates at a closing XOR-transition $t_{XOR}$ with one high-marked and one unmarked preplace. In the other case $t^{j,fin}$ is the high-mode of a closing

AND-transition and its preplace different from $p^j$ is unmarked. This proves the assumption valid while entering the loop for the next run with index $j+1$.

ii) Termination: First we prove, that the case $q \in N^k$ in the loop body will never happen for an index $k < j$. Assume on the contrary a first index $j \geq 1$ with

$$q \in U := \bigcup_{k=0}^{j-1} N^k \subset N^{flat,high}.$$

The case $j = 1$ with $q \in N^0$ cannot happen, because the P-component $N^0$ already contains $p^0$ and both places $p^0, q \in N^{flat.high}$ are preplaces of the same transition. If $j \geq 2$ we denote by $D \subset U$ the set of all unmarked places of $U$. This set is not empty, because $q \in D$. We claim $pre(D) \subset post(D)$, i.e. for every place $s^{high} \in D$ every pretransition

$$t^{high} \in pre( s^{high} ) \subset N^{flat,high}$$

has a preplace in $D$. For the proof denote by

$$s := high^{-1}( col(s^{high})) \in N \text{ and } t := high^{-1}( col( t^{high} )) \in pre( s ) \subset N$$

respectively the place belonging to $s^{high}$ and the transition belonging to $t^{high}$ from the corresponding BP-graph $N$. According to Lemma 4.1 we distinguish the following cases for $t$:

- For an opening transition $t$ every preplace of $t \in N$ is unmarked, hence also each preplace of $t^{high}$.
- A closing XOR-transition $t$ cannot have a high-marked preplace, because in that case the algorithm had already terminated during a previous loop. Hence again every preplace of $t^{high}$ is unmarked.
- If a closing AND-transition $t$ has a high-marked preplace, the other preplace is unmarked and already belongs to $U$ by construction.

In every case $t^{high}$ has a preplace from $D$. Because the starting place $s^{high} \in D$ was arbitrary, we have proved that $D$ is a non-empty, unmarked siphon of the flat high-system. This fact contradicts the liveness of the high-system and proves $q \notin U$. Hence, during the loop with index $j$ the union $\bigcup_{k=0}^{j} N^k$ increases at least by the place $q$. After a finite number $n$ of iterations the termination condition is satisfied and the algorithm terminates.

iii) Correctness of the output: By construction the closing XOR-transition $t_{XOR} = t_0$ is the initial transition of the AND/XOR-chain $( t_i, N_i, p_i )_{i=0,...,n}$ with $t_{AND} = t_n$, q. e. d.

## 4.5 **Lemma** *(Deadlock-configuration)*

For a dead BP-system it is impossible, that its skeleton and its high-system are safe and live and its high-system has no frozen tokens.

**Proof.** For an indirect proof we consider a dead BP-system $BPS = (N, \mu)$ with a safe and live skeleton and a safe and live high-system without frozen tokens.

i) Existence of a minimal deadlock-configuration of positive size: The dead BP-system $BPS$ has a non-empty set of closing AND-transitions in deadlock by Lemma 4.1 and the images of their final transitions are the only transitions activated in the skeleton. Algorithm 4.4 completes each closing AND-transition in deadlock to an AND/XOR-chain. Enumerate all AND/XOR-chains of $BPS$ as $Ch_{i,AND/XOR}$, $i = 0, ..., r-1$. Because the skeleton is live, each initial transition of an AND/XOR-chain can be reached from the final transition of the same or another AND/XOR-chain by an unmarked path. After possibly renumbering a subset of AND/XOR-chains we obtain a deadlock-configuration. Hence $BPS$ has also a minimal deadlock-configuration

$$\left( Ch_{i,AND/XOR}, \beta_i \right)_{i=0,...,m-1}, \; m \geq 1.$$

ii) Selecting basic circuits: For every index $i = 0, ..., m-1$ and AND/XOR-chain

$$Ch_{i,AND/XOR} = \left( t_{i,k}, N_{i,k}, p_{i,k} \right)_{k=0,...,n}$$

from $t_{i,XOR} = t_{i,0}$ to $t_{i,AND} = t_{i,n}$ we select a basic circuit $BK_i$ passing through the high-marked preplace of $t_{i,XOR}$. It does not pass through any of the deadlock-transitions $t_{j,AND}$, $j = 0, ..., m-1$, because both of their preplaces are marked.

Each $BK_i$ determines in the flat high-system a basic circuit $BK_i^{high} \subset BPS^{flat,high}$ with

$$col\left(BK_i^{high}\right) = high\left(BK_i\right) \subset N^{high}.$$

Claim: For every pair $i, j = 0, ..., m-1$ holds

$$BK_i \cap \beta_j = \begin{cases} \emptyset & i \neq j+1 \\ \left\{ t_{i,XOR}, post\left(t_{i,XOR}\right) \right\} & i = j+1 \end{cases}$$

For the proof assume on the contrary the existence of a node $x \in BK_i \cap \beta_j$, $x \notin \left\{ t_{i,XOR}, post\left(t_{i,XOR}\right) \right\}$. In case $i = j+1$ we obtain an unmarked circuit by composing the segment of $\beta_j$ from $x$ to $t_{i,XOR}$ with the segment of $BK_i$ from $t_{i,XOR}$ to $x$, which is a contradiction to the liveness of the skeleton. In case $i \neq j+1$ we obtain an unmarked path from $t_{i-1,AND}$ to $post\left(t_{j+1,XOR}\right)$ by composing three single paths: $\beta_{i-1}$, secondly the segment of $BK_i$ from $post\left(t_{i,XOR}\right)$ to $x$ and as third the segment of $\beta_j$ from $x$ to $post\left(t_{j+1,XOR}\right)$. Connecting $CH_{i-1,AND/XOR}$ and $CH_{j+1,AND/XOR}$ by the resulting path and skipping all AND/XOR-chains

$$CH_{k,AND/XOR} \text{ with } i \le k \le j$$

produces a deadlock-configuration of smaller size than the original, minimal one.

iii) Selecting a path from the initial to the final transition of each AND/XOR-chain: By Corollary 1.2 there exists a T-component $N_{i,T} \supset BK_i^{high}$, and by Lemma 4.3 it passes through all transitions $t_{i,k}^{high}$, $k \in \{1,...,n\}$. We define a path $\alpha_i \subset N_{i,T}$ from the postplace of the activated transition $t_{i,XOR}^{high}$ to $t_{i,AND}^{high}$ as follows: For every index $k \in \{1,...,n\}$ the intersection $K_{i,k} := N_{i,k} \cap N_{i,T}$ with the basic component $N_{i,k}$ is an elementary circuit according to Corollary 1.6. It passes through the marked preplace $p_{i,k}$ of $t_{i,k}^{high}$ and through the unmarked preplace of $t_{i,k+1}^{high}$. We denote by $B_{i,1} \subset K_{i,1}$ its segment from the postplace of $t_{i,0}^{high}$ to $t_{i,1}^{high}$ and for $2 \le k \le n$ by $B_{i,k} \subset K_{i,k}$ its segment from $t_{i,k}^{high}$ to $t_{i,k+1}^{high}$. We define the composed path

$$\alpha_i := B_{i,n} \circ ... \circ B_{i,1}.$$

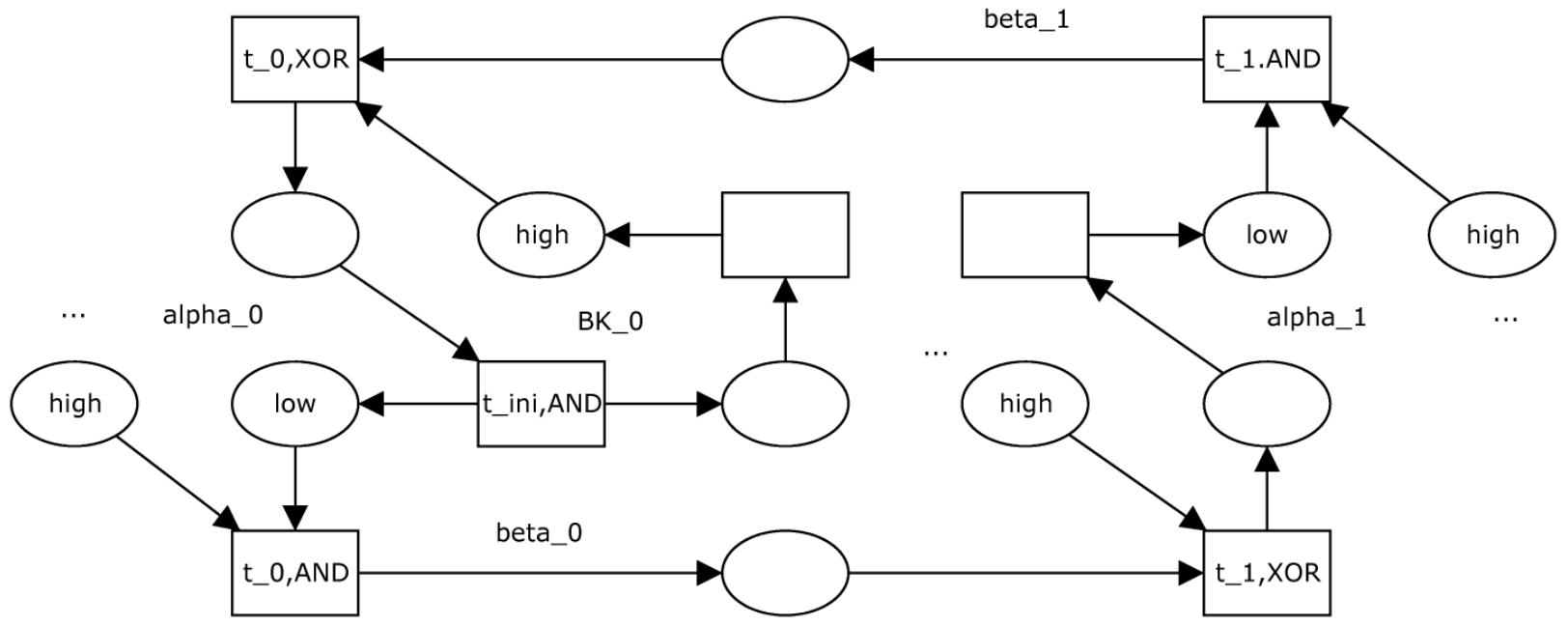

**Fig. 3**. Deadlock-configuration of size $m = 2$

iv) Derivation of a TP-handle: We select an arbitrary, but fixed index from $\{0,...,m-1\}$, e.g., the index $0$. Let $i \in \{0,...,m-1\}$ be maximal with $t_{i,AND}^{high} \in N_{0,T}$. Such an index exists, because at least $t_{0,AND}^{high} \in N_{0,T}$. Let $B \subset N_{0,T}$ be a bridge from the elementary circuit $BK^{high} := BK_0^{high}$ to $t_{i,AND}^{high} \in N_{0,T} - BK^{high}$. As a path within a T-component $B$ starts with a transition $t_{ini}$. It is the high-mode of an AND-transition $t_{ini,AND}$, cf. Fig. 3. Claim: The path

$$H := \beta_{m-1}^{high} \circ \alpha_{m-1} \circ ... \alpha_{i+1} \circ \beta_i^{high} \circ B \subset N^{flat,high}$$

can be altered to a TP-handle $\overline{H}$ on $BK^{high}$, keeping fixed its ends $t_{ini}$ and $p_{fin} := post\left(t_{0,XOR}{}^{high}\right)$. In case $H$ intersects $BK^{high}$ in between, we have

$$\left(BK^{high} \cap H\right) \underset{\neq}{\supset} \left\{t_{ini}, p_{fin}\right\}.$$

Then there exists due to part ii) at least one index $j$ with

$$i+1 \leq j \leq m-1 \text{ and } \left(BK^{high} \cap H\right) \cap \alpha_j \neq \emptyset .$$

We consider all indices $j$ with this property. For each corresponding AND/XOR-chain

$$Ch_{j,AND/XOR} = \left(t_{j,k}, N_{j,k}, p_{j,k}\right)_{k=0,\dots,n}$$

holds $t_{j,AND}{}^{high} \notin N_{0,T}$ according to our choice of $i$, hence by Lemma 4.3 also $t_{j,k}{}^{high} \notin N_{0,T}$ for all $k = 0,\dots,n$. Theorem 1.1 proves the existence of a minimal occurrence sequence $\sigma$, activating $N_{0,T}$, but firing no transition from $cl\left(N_{0,T}\right)$, in particular no transition from $cl\left(BK^{high}\right)$. Hence $\sigma$ neither fires a transition from $BK^{high}$ nor a transition removing a token from $BK^{high}$. Therefore the initial token on the preplace of $t_{0,XOR}{}^{high}$ remains unchanged during firing $\sigma$, alike any additional token possibly generated by $\sigma$ on a place of $BK^{high}$. On the other hand $\sigma$ fires every transition $t_{j,k}{}^{high}$, $k = 0,\dots,n$, because $t_{j,k}{}^{high}$ has a marked preplace, but does not belong to $N_{0,T}$ itself. In order to fire $t_{j,k}{}^{high}$ the occurrence sequence has to move the well determined token of the basic component $N_{j,k}$ to the unmarked preplace of $t_{j,k}{}^{high}$, in descending order for $k = n,\dots,1$. By the reasoning above this token has to avoid all places of $BK^{high}$. Hence the token flow defines a path $\overline{\alpha}_j$ from $t_{j,XOR}{}^{high}$ to $t_{j,AND}{}^{high}$ with $\overline{\alpha}_j \cap BK^{high} = \emptyset$. For every index $j$ with $i+1 \leq j \leq m-1$ and $\left(BK^{high} \cap H\right) \cap \alpha_j \neq \emptyset$ we replace in $H$ the path $\alpha_j$ by $\overline{\alpha}_j$. The resulting path $\overline{H}$ is disjoint from $BK^{high}$ with exception of its ends, hence a TP-handle on $BK^{high}$ after shortening to an elementary path. This proves the claim.

By Theorem 1.7 the existence of a TP-handle on the elementary circuit $BK^{high}$ contradicts the well-formedness of the high-net and finishes the proof of the Lemma, q. e. d.

**Note.** The transitions of the BP-systems from the rest of this chapter are not necessarily binary.

## 4.6 Theorem *(Liveness and safeness of BP-systems)*

A BP-system is safe and live with respect to all its high-modes if and only if its skeleton is safe and live and its high-system is safe and live without frozen tokens.

**Proof.** One direction is Theorem 3.6. To prove the other direction: The safeness of the skeleton implies the safeness of the BP-system according to Lemma 3.1. To prove its liveness with respect to all high-modes it suffices according to Corollary 3.7 to prove its high-liveness. Herefore it suffices according to Corollary 3.4 to exclude that a reachable marking is dead. Assume on the contrary that $BPS$ has a reachable dead marking. Without loss of generality already $BPS$ is dead. Now Lemma 4.5 provides a contradiction, q. e. d.

The statement of the following Corollary 4.7 is due to Genrich and Thiagarajan ([GT1984], Theor. 4.10).

## 4.7 Corollary *(Full reachability class)*

A BP-system $(N, \mu_0)$ is safe and live with respect to all its high-modes, iff $(N, \mu)$ is safe and live with respect to all its high-modes for every marking $\mu \in [\ \mu_0\ ]$ from the full reachability class of $\mu_0$.

**Proof.** Only one direction needs an explicit proof: We assume, that $\mu_0$ is reachable in $(N, \mu)$ and that $(N, \mu_0)$ is safe and live with respect to all high-modes. We have to prove, that also $(N, \mu)$ is safe and live with respect to all high-modes: The morphisms

$$(N, \mu) \xrightarrow{skel} \left(N^{skel}, \mu^{skel}\right) \text{ and } \left(N^{flat}, \mu^{flat}\right) \xrightarrow{high} \left(N^{flat,high}, \mu^{flat,high}\right)$$

imply, that $\mu_0^{skel}$ is reachable in $\left(N^{skel}, \mu^{skel}\right)$ and $\mu_0^{flat,high}$ in $\left(N^{flat,high}, \mu^{flat,high}\right)$. The P-coverability theorem for a well-formed free-choice net ([DE1995], Theor. 5.6) implies, that every marking from the full reachability class of a safe and live marking is safe and live itself. Hence $\left(N^{skel}, \mu^{skel}\right)$ as well as $\left(N^{flat,high}, \mu^{flat,high}\right)$ is safe and live. By Theorem 3.6 and Lemma 1.5 the net $N^{flat,high}$ is structural free from blocking. Now Theorem 4.6 implies, that $(N, \mu)$ is safe and live with respect to all its high-modes, q. e. d.

The following Lemma 4.8 prepares the proof of Theorem 4.9. It proves, that reverse firing in the high-system of a BP-system lifts to the BP-system itself.

## 4.8 Lemma *(Reverse firing)*

Consider a BP-system $(BPG, \mu)$, which is safe and live with respect to all its high-modes. If a marking $\mu_0^{high}$ of the high-net activates an occurrence sequence $\sigma^{high}$ with fires according to

$$\mu_0^{high} \xrightarrow{\sigma^{high}} high(\mu),$$

then there exists a marking $\mu_0$ of $BPG$ and an occurrence sequence $\sigma$ of $(BPG, \mu_0)$, such that the following diagram commutes

$$\begin{array}{ccc} \mu_0 & \xrightarrow{\sigma} & \mu \\ high \downarrow & & \downarrow high \\ \mu_0^{high} & \xrightarrow{\sigma^{high}} & high(\mu) \end{array}$$

**Proof.** We denote by $N := BPG^{high}$ the high-net and set $FCS := \left(N, \mu_0^{high}\right)$. Due to Lemma 1.5 also $FCS$ has no frozen tokens. Without loss of generality we may assume that $\sigma^{high}$ is a single transition $\sigma^{high} = t^{high}$. There exists a well-determined firing element $(t, b) \in BPG$ with $high(t,b) = col\left(t^{high}\right)$. For the token changes $\Delta\mu\left(\sigma^{high}\right)$ due to the firing of $\sigma^{high}$ and $\Delta\mu(\sigma)$ due to the firing of $\sigma := (t, b)$ in $BPG$ holds

$$high(\ \Delta\mu(\sigma)\ ) = col\left(\ \Delta\mu\left(\sigma^{high}\right)\ \right).$$

i) In case of a closing AND-transition or an arbitrary opening transition $t$ all of its postplaces are high-marked at $\mu$ and we have $\mu - \Delta\mu(\sigma) > 0$. Hence

$$\mu_0 := \mu - \Delta\mu(\sigma)$$

is a marking of $BPG$ and provides a lift with the necessary properties.

ii) The remaining case considers an opening XOR-transition $t = t_{XOR}$. Denote by $p_2$ the preplace of $t_{XOR}$, by $p_1$ the postplace of $t_{XOR}$, which is high-marked at $\mu$. The other postplace $p_3$ of $t_{XOR}$ possibly lacks a low-token. Hence not necessarily $\mu - \Delta\mu(\sigma) > 0$ and this expression may fail to define a marking of $BPG$. If $p_3$ lacks a low-token at $\mu$, we have to fire the low-system in reverse direction until reaching a low-token at $p_3$. This can be done by firing the skeleton in reverse direction from the marking $skel(\mu)$ and making sure, that it lifts to the reverse of the low-system $BPS^{low}$. The skeleton $BPS^{skel}$ is a safe and live T-system. By reversing the orientation of its arcs - but keeping transitions, places and markings – we obtain the reverse of the skeleton, which is a safe and live T-system too. We select a minimal occurrence sequence of the reverse skeleton, which activates the transition $skel(t_{XOR})$ for reverse firing. To show that it lifts to the reverse of $BPS^{low}$, we have to exclude the existence of an elementary path in $BPS$, which starts at $p_3$ and contains a single token, which is a high-token at its final place. For an indirect proof we assume the existence of such a path $\alpha$. First we claim, that no transition from the postset $post(post(t_{XOR}))$ is a closing AND-transition: Otherwise we consider the high-net. A closing AND-transition

$$t_{AND} \in post(post(t_{XOR})) \subset BPG$$

induces in the high-net a P-component $N_P$, passing through $t_{AND}{}^{high}$ but not through the preplace $p_2{}^{high}$, as well as a T-component $N_T$, passing through $p_2{}^{high}$ but not through $t_{AND}{}^{high}$. The intersection $BK := N_T \cap N_P$ is an elementary circuit according to Corollary 1.6. We compose three pathes: A bridge from $BK$ to $t_{AND}{}^{high}$ within $N_T$, the direct elementary path from $p_2{}^{high}$ to $t_{AND}{}^{high}$ and as third a bridge from $t_{AND}{}^{high}$ to $BK$ within $N_P$. The composed path is a TP-handle at $BK$. Due to Theorem 1.7 such a handle contradicts the well-formedness of the high-net. This contradiction proves the claim.

Because $FCS$ has no frozen tokens, we can block any cluster of the high-net. We consider the two clusters $cl\left(p_i{}^{high}\right), i = 1, 2$, of the high-net and denote by

$$\mu^{high}{}_{i,block}\ , i = 1, 2,$$

their blocking markings in $FCS$. They are uniquely determined by Lemma 1.9. There exists a minimal occurrence sequence $\sigma^{high}{}_{1,block}$, which fires according to

$$\mu^{high} \xrightarrow{\sigma^{high}{}_{1,block}} \mu^{high}{}_{1,block}$$

The occurrence sequence $\sigma^{high}{}_{1,block}$ lifts according to Corollary 3.5 to an occurrence sequence $\sigma_1$ of $(BPG, \mu)$. By possibly firing the low-system $\sigma_1$ extends to an occurrence sequence $\sigma_{1,block}$ of $(BPG, \mu)$, which fires according to

$$\mu \xrightarrow{\sigma_{1,block}} \mu_1,$$

such that $high(\mu_1) = \mu^{high}{}_{1,block}$ is the blocking marking of $cl\left(p_1{}^{high}\right)$ from the high-system and $skel(\mu_1)$ is the blocking marking of $cl(skel(p_1))$ from the skeleton. Because the skeleton is a T-system and $t_{XOR}$ does not fire neither in $\sigma_1$ nor in $\sigma_{1,block}$, the path $\alpha$ extends to a path $\tilde{\alpha}$ from $skel(p_3)$ to a place $q^{skel}$, such that at $skel(\mu_1)$ the place $q^{skel}$ is marked and $\tilde{\alpha}$ contains no other tokens.

Now we select a minimal occurrence sequence $\sigma^{high}{}_{2,block}$ of $FCS$, which fires according to

$$\mu^{high}{}_{1,block} \xrightarrow{\sigma^{high}{}_{2,block}} \mu^{high}{}_{2,block}\ .$$

Again the occurrence sequence $\sigma^{high}{}_{2,block}$ lifts to an occurrence sequence $\sigma_2$ of $(BPG, \mu)$, which fires according to

$$\mu_1 \xrightarrow{\sigma_2} \mu_2\ .$$

At $\mu_2$ we fire that high-mode of $t_{XOR}$, which creates a marking $\mu_3$ of $(BPG, \mu)$ that high-marks $p_1$ and low-marks $p_3$. Because $post(post(t_{XOR}))$ does not contain a closing

AND-transition, the marking $high(\mu_3) = \mu^{high}{}_{1,block}$ is the blocking marking of $cl\left(p_1^{high}\right)$ from the high-system. After possibly firing the low-system from $\mu_3$ we obtain a reachable marking $\mu_4$ of $(BPG, \mu)$, such that $high(\mu_4) = \mu^{high}{}_{1,block}$ and $skel(\mu_4)$ is the blocking marking of $cl(skel(p_1))$ from the skeleton. But both blocking markings $skel(\mu_1)$ and $skel(\mu_4)$ differ: At $skel(\mu_4)$ the path $\tilde{\alpha}$ contains a second token originating from the low-token, which marks $p_3$ at $\mu_3$. This contradicts the uniqueness of blocking markings of a fixed cluster from the skeleton and completes the proof ot the lemma, q e. d.

The next Theorem 4.9 answers in the positive a question of Jörg Desel[2].

## 4.9 Theorem *(Extending free-choice systems to live BP-systems)*

Any restricted free-choice system, which is safe and live without frozen tokens, is the high-system of a BP-system, which is safe and live with respect to all its high-modes.

**Proof.** We denote by $FCS = \left(N, \mu_0^{high}\right)$ the given free-choice system. For the proof we may assume that all transitions of $N$ are unary or binary - and similar for places of $N$.

i) Catching all high-tokens within a T-component: We choose a T-component $N_T$ of $N$. According to Theorem 1.1 there exists a reachable marking $\mu_1^{high}$ of $FCS$, which activates $N_T$. The component $N_T$ contains all tokens of $\mu_1^{high}$, because $FCS$ has no frozen tokens, The component is covered by a family of basic circuits and each elementary circuit of $N_T$ is marked at $\mu_1^{high}$. Denote by

$$\Gamma := \{\, \gamma \subset N_T : \gamma \ elementary \ circuit \,\}$$

the set of all elementary circuits in $N_T$.

ii) Adding low-tokens: The restricted free-choice net $N$ extends to a unique BP-graph $BPG$ with high-net $BPG^{flat,high} = N$ : The BP-graph $BPG$ has an XOR-transition with adjacent places for every branched place of $N$ and an AND-transition for every branched transition of $N$. Now we follow the iterative procedure in the proof of Genrichs Theorem ([DE1995], Theor. 3.20). Using the morphisms

$$N \xrightarrow{col} BPG^{high},\ BPG \xrightarrow{high} BPG^{high},\ BPG \xrightarrow{skel} BPG^{skel}$$

we shall produce a certain save and life marking $\mu^{skel}$ of the skeleton $BPG^{skel}$ without changing the token content of any elementary circuit

[2] Personal communication 15.9.2006.

$$\gamma^{skel} := skel\left( high^{-1}\left( col(\gamma) \right) \right), \gamma \in \Gamma .$$

These circuits cover

$$N_T{}^{skel} := skel\left( high^{-1}\left( col(N_T) \right) \right),$$

the subnet of $BPG^{skel}$ corresponding to $N_T \subset BPG^{flat,high}$. To start the iteration we lift the marking $\mu_1{}^{high}$ from $BPG^{flat,high}$ to the well-defined marking $\mu_{1,h}$ of high-tokens on $BPG$ with $high\left(\mu_{1,h}\right) = col\left(\mu_1{}^{high}\right)$. We extend $skel(\mu_{1,h})$ to a live marking $\mu_1{}^{skel}$ of $BPG^{skel}$ by adding a token to each place from $BPG^{skel} - N_T{}^{skel}$. The marking does not change the token content of any elementary circuit $\gamma^{skel}$, $\gamma \in \Gamma$. If the marking $\mu_1{}^{skel}$ is not safe already, there exists a reachable marking of $\left(BPG^{skel}, \mu_1{}^{skel}\right)$, which marks a certain place of $BPG^{skel}$ with two or more tokens. This place must belong to $BPG^{skel} - N_T{}^{skel}$, because $N_T{}^{skel}$ is covered by basic circuits. After removing all but one token from the place in question the resulting marking is still live, but the token content has decreased for at least one circuit not contained in $N_T{}^{skel}$. We iterate this step until the resulting marking $\mu^{skel}$ of $BPG^{skel}$ is also safe.

iii) Extending a certain reachable marking of $FCS$ to $BPG$: We lift the restriction $\mu^{skel} \mid N_T{}^{skel}$ to the well-defined marking $\mu_h$ of high-tokens on $BPG$ with $skel(\mu_h) = \mu^{skel} \mid N_T{}^{skel}$. There exists a well-defined marking $\mu^{high}$ on $N_T$ with $high(\mu_h) = col\left( \mu^{high} \right)$. The two markings $\mu_1{}^{high} \mid N_T$ and $\mu^{high}$ of the T-net $N_T$ agree on all P-flows of $N_T$, because they have the same token content on all elementary circuits $\gamma \in \Gamma$. Hence the marking $\mu^{high}$ is a reachable marking of $\left(N_T, \mu_1{}^{high} \mid N_T\right)$ according to the Reachability Theorem for live T-systems ([1995], Theor. 3.21). Because $N_T$ is even a T-component of $N$, the marking $\mu^{high}$ is also reachable in $\left(N, \mu_1{}^{high}\right)$ as well as in the original system $FSC$. Analogously we lift the restriction $\mu^{skel} \mid BPG^{skel} - N_T{}^{skel}$ to the well-defined marking $\mu_l$ of low-tokens on $BPG$ with $skel(\mu_l) = \mu^{skel} \mid BPG^{skel} - N_T{}^{skel}$. The combined marking

$$\mu := \mu_h + \mu_l$$

defines the BP-system $(BPG, \mu)$. Its high-system $\left(N, \mu^{high}\right)$ is safe and live without frozen tokens, and its skeleton $\left(BPG^{skel}, \mu^{skel}\right)$ is safe and live. Hence $(BPG, \mu)$ is safe and live with respect to all its high-modes according to Theorem 4.6.

To complete the proof of the theorem we apply Lemma 4.8. It implies the existence of a marking $\mu_0$ of $BPG$, such that $FCS$ is the high-system of the BP-system $BPS := (BPG, \mu_0)$, which is safe and live with respect to all high-modes according to Lemma 4.7, q. e. d.

## 4.10 Example *(Obstructions due to frozen tokens)*

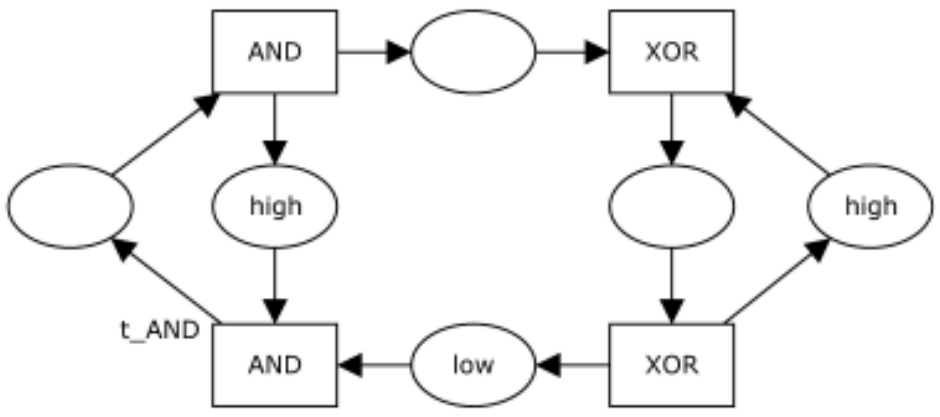

**Fig. 4**. BP-system with frozen token in the high-system

The BP-system from Fig. 4 is dead. Its skeleton and its high-system are safe and live. But the high-system has a frozen token resulting from the high-marked preplace of $t_{AND}$. The example shows, that the assumptions in Lemma 4.5 and Theorem 4.6 concerning the absence of frozen tokens are necesssary.

## 4.11 Remark *(Reduction and synthesis)*

Genrich and Thiagarajan developed an algorithm for reducing a BP-system ([GT1984], Theor. 6.19): A safe BP-system is live with respect to all high-modes, iff it is reducible to any of two trivial normal forms. According to Theorem 4.6 reduction and synthesis of those BP-systems, which are safe and live with respect to all high-modes, is a special case of the analogous problem for safe and live free-choice systems without frozen tokens. This problem has already been solved by Desel [Des1992], his algorithm has polynomial time complexity.

# 5 Perspectives

In a previous paper [LSW1998] we used BP-systems to provide a widespread language from business administration with a Petri net semantics; namely, the language of Event-triggered Process Chains (EPC). We proposed the translation of an EPC into a Boolean loop tree, which is composed of BP-systems. Now Theorem 4.6 reduces the verification of these EPCs to the verification of liveness, safeness and the absence of frozen tokens of restricted free-choice systems. The latter verification has polynomial time complexity. We will investigate elsewhere the consequences for the domain of business administration.

The present paper exemplified how to study Petri nets using morphisms. Essential properties of the BP-system $BPS$ in the domain of the two morphisms

$$BPS \xrightarrow{high} BPS^{high} \text{ and } BPS \xrightarrow{skel} BPS^{skel}$$

result from the corresponding properties of the two Petri nets $BPS^{high}$ and $BPS^{skel}$ in the range of the morphisms. Moreover, the coloured Petri net $BPS$ is an extension of the free-choice system $BPS^{high}$ by the T-system $BPS^{low}$ according to

$$BPS^{low} \xrightarrow{low} BPS \xrightarrow{high} BPS^{high}.$$

It seems worthwhile to investigate the basic mathematical concept of a morphism in the context of Petri nets and to study the topological and algebraic aspects of a Petri net morphism.